\documentclass[english]{article}
\usepackage{lmodern}
\usepackage[T1]{fontenc}
\usepackage[latin9]{inputenc}
\usepackage{amsmath}
\usepackage{amssymb}
\usepackage{graphicx}
\usepackage{babel}
\usepackage{geometry}
\usepackage{authblk}

\begin{document}

\title{Cosmological scenario based on the particle creation and holographic
equipartition}

\author[a]{Fei-Quan Tu \thanks{Corresponding author: fqtuzju@foxmail.com}}
\author[b]{Yi-Xin Chen}
\author[a]{Qi-Hong Huang}
\affil[a]{School of Physics and Electronic Science, Zunyi Normal University, Zunyi 563006, China}
\affil[b]{Zhejiang Institute of Modern Physics, Zhejiang University, Hangzhou 310027, China}


\renewcommand*{\Affilfont}{\small\it} 
\renewcommand\Authands{ and } 
\date{} 

\maketitle
\begin{abstract}
We propose a cosmological scenario which describes the evolution history
of the universe  based on the particle creation
and holographic equipartition. The model attempts to solve the inflation
of the early universe and the accelerated expansion of the present
universe without introducing the dark energy from
the perspective of thermodynamics. Throughout the evolution of the
universe, we assume that the universe always creates particles in
some way and holographic equipartition is always satisfied. Further,
we choose that the creation rate of particles is proportional
to $H^{2}$ in the early universe and to $H$ in the present and late
universe, where $H$ is the Hubble parameter. Then we obtain the solutions
$a(t)\propto e^{\alpha t/3}$ and $a(t)\propto t^{1/2}$ for the early universe
and the solutions $a(t)\propto t^{\delta}$ and $a(t)\propto e^{Ht}$
for the present and late universe, where $\alpha$ and $\delta$ are the parameters.
Finally, we obtain and analyze two important thermodynamic properties for the present
model.
\end{abstract}

\section*{1. Introduction}

The accelerated expansion of the universe have been confirmed by the
probe of Type Ia supernovae\cite{key-1,key-2}, cosmic microwave background
radiation\cite{key-3,key-4} and baryon acoustic oscillations\cite{key-5,key-6}
since the 1990s. In order to explain the phenomenon, various models
such as the lambda cold dark matter ($\Lambda\mathrm{CDM}$) model\cite{key-7},
$\Lambda(t)\,\mathrm{CDM}$ (i.e. the dark energy varies with time
$t$) model\cite{key-8,key-9} and particle creation model\cite{key-10,key-11}
have been proposed. Among these models, the particle creation model
has some following advantages\cite{key-11,key-12}: (i) There exists
a description of the non-equilibrium thermodynamics because the process of particle creation
is an irreversible process accompanied by the entropy production\cite{key-13,key-14}.
(ii) It contains only a single free parameter, namely the particle creation rate.
Such a one-parameter model is preferred by the statistical
Bayesian analysis\cite{key-15}.

In the past 30 years, great progress has been made in the study of
gravity from the viewpoint of thermodynamics. In 1995, Jacobson showed
that the Einstein field equation can be derived based on the Clausius
relation and entropy-area relation\cite{key-16}. Then he and his
collaborators investigated the properties of the non-equilibrium thermodynamics
of the spacetime by introducing the term of entropy generation\cite{key-17}.
On the other hand, Padmanabhan derived the important thermodynamic
relation $S=2\beta E$\cite{key-18} and the first law of thermodynamics\cite{key-19}
consistent with the usual form of thermodynamics from a similar but
different thermodynamic perspective, where $S$ is the entropy of
the horizon, $\beta=1/(kT)$ in which $k$ is the Boltzmann constant
and $T$ is the temperature of the horizon, and $E$ represents the
active gravitational mass. In addition, there exists a simple correspondence
between the surface term $L_{sur}$ and bulk term $L_{bulk}$ when
the Lagrangian of gravity $L$ is decomposed into $L=L_{sur}+L_{bulk}$\cite{key-20,key-21},
which shows that the gravitation action is holographic because the
same information is coded in the surface term and bulk term. These results
indicate that there is a deep connection between gravitational dynamics
and horizon thermodynamics.

In particular, Padmanabhan proposed that the spacetime is composed
of the microscopic degree of freedom and the de Sitter universe satisfies
the holographic equipartition (an especial holographic principle)\cite{key-22,key-23}.
Further, he thought that our universe is asymptotically de Sitter
and derived the evolution equations of the universe based on the difference
between the number of degrees of freedom on the surface and the number of degrees of freedom in the
bulk. Subsequently, this model was extensively studied\cite{key-24,key-25,key-26,key-27,key-28,key-29,key-30,key-31,key-32}.
Therefore, there is a solid physical foundation to study the evolution
of the universe from the perspective of holographic equipartition.

In this paper, we propose a cosmological scenario which describes
the evolution history of the universe  based on the particle creation and
holographic equipartition. Throughout the evolution of the universe,
we assume that the universe always creates particles in some way and
holographic equipartition is always satisfied. In our scenario,
the creation rate of the radiation is $\Gamma=\alpha H^{2}$ in the
early universe and the creation rate of the pressureless matter is $\Gamma=\alpha H$
in the present and late universe, where $\alpha$ is a positive parameter
and $H$ is the Hubble parameter. The whole evolution history of
the universe may be explained as follows. The universe starts from
an unstable de Sitter universe ($a(t)\propto e^{\alpha t/3}$) and evolves into a
standard radiation stage ($a(t)\propto t^{1/2}$) due to the creation
of radiation. With the expansion of the universe, the pressureless matter whose
creation rate is $\Gamma=\alpha H$ begins to dominate the universe.
The negative creation pressure of the matter accelerates the expansion
of the universe and drives the present accelerated universe ($a(t)\propto t^{\delta}\,(\delta>1)$) to the
de Sitter universe ($a(t)\propto e^{Ht}$).

The paper is organized as follows. In section 2, we introduce the
non-equilibrium thermodynamics which describes the particle creation.
In section 3, we obtain the energy of the universe enclosed by the
Hubble horizon by using the laws of energy equipartition and holographic
equipartition. Next we obtain the laws of evolution for the early
universe and the present and late universe by choosing the
suitable creation rates of particles in section 4.
In section 5, we investigate the thermodynamical properties
 for this model. The summary is made
in the last section. We use units $\ensuremath{c=\hbar=1}$.

\section*{2. Particle creation in cosmology }

As we know, the first law of thermodynamics in a close system which
only does the volume work can be expressed as
\begin{equation}
dE=\delta Q-pdV,
\end{equation}
where $\delta Q$ is the amount of heat exchanged by the system,
$p$ is the pressure, $dE$ is the change of
internal energy and $dV$ is the change of volume. When the thermodynamic
process is reversible, the amount of heat exchanged by the system
can be expressed as $\delta Q=TdS$ where $T$ is the temperature
and $dS$ is the entropy change.

For an adiabatic open system where the particle number is not conserved
due to the particle creation, the first law of thermodynamics is written
as\cite{key-10}
\begin{equation}
d(\rho V)+pdV=\frac{h}{n}d(nV),
\end{equation}
where $n$ is the particle number per unit volume, $h=\rho+p$
is the enthalpy with $\rho$ being the energy density.

To calculate the creation pressure due to the particle creation, we
consider the universe as a sphere with the radius $a(t)$. Thus the volume
is $V=\frac{4\pi}{3}a^{3}(t)$ and the right term of
Eq.(2) is expressed as
\begin{equation}
\frac{h}{n}d(nV)=(\rho+p)V\left(3H+\frac{\dot{n}}{n}\right)dt,
\end{equation}
where $H=\frac{\dot{a}}{a}$ is the Hubble parameter. Inserting Eq.(3)
and the relation $\dot{n}+3Hn=n\Gamma$ \cite{key-10,key-33,key-34,key-35} into Eq.(2)
where $\Gamma$ is the particle
creation rate, we obtain
\begin{equation}
\dot{\rho}+3H(\rho+p+p_{e})=0,
\end{equation}
where the effect of particle creation is expressed as an extra
pressure
\begin{equation}
p_{e}=-\frac{(\rho+p)\Gamma}{3H}.
\end{equation}
Thus we can deem that Eq.(4) is the modified continuity equation due to the particle creation.

Particle creation in the universe generates the entropy whose change is $dS=\frac{s}{n}d(nV)$ where $s=S/V$\cite{key-10},
which provides an explanation for the origin of cosmological entropy.
In addition, the negative pressure caused by the particle creation can explain the current accelerated expansion.
In fact, introducing bulk viscous fluids to study the
accelerated expansion of the universe is also due to the negative
pressure effect of viscous fluids\cite{key-36,key-37,key-38,key-39}.

\section*{3. Energy equipartition and holographic equipartition}

Research in recent decades has uncovered an important fact that the
spacetime can be heated up like the matter. In fact, Unruh discovered
that an observer can measure a temperature $T=\frac{\kappa}{2\pi}$
when he accelerates through the inertial vacuum with a proper acceleration
$\kappa$ in 1976\cite{key-40}. Furthermore, the temperature
measured by the accelerated observer in the vacuum is as real
as that measured by an inertial thermometer in the ordinary matter\cite{key-41}.
Hence, it can be concluded that the spacetime, just as the ordinary
matter, is made up of the microscopic degrees of freedom. In the description
of the evolution of the universe based on the emergent perspective
of gravity, the energy equipartition law
\begin{equation}
E=\frac{1}{2}N_{bulk}kT
\end{equation}
holds, where $E$ is the energy in the volume $V$, $N_{bulk}$ is the
number of degrees of freedom in the volume $V$ and $T$ is the local
acceleration temperature.

In Ref.\cite{key-22}, Padmanabhan thought that our universe is asymptotically
de Sitter rather than exactly de Sitter and the expansion of the universe
is being driven by the difference ($N_{sur}-N_{bulk}$) where $N_{sur}$
is the number of surface degrees of freedom on the horizon. However,
we assume that the universe obeys always holographic equipartition due to the
particle creation in this paper. Therefore the number of bulk degrees of freedom
enclosed by the horizon is
\begin{equation}
N_{bulk}=N_{sur}.
\end{equation}
In addition, the number of surface degrees of freedom can be expressed
as\cite{key-42}
\begin{equation}
N_{sur}=\frac{A}{L_{p}^{2}},
\end{equation}
where $A$ is the area of the horizon. So
the total energy in the volume $V$ is
\begin{equation}
E=\frac{A}{2L_{p}^{2}}kT.
\end{equation}

Now let us consider an universe enclosed by the Hubble horizon which
is a sphere with the radius $H^{-1}$. Thus, the volume, the temperature
and the area of the Hubble horizon can be written as $V=\frac{4\pi}{3H^{3}},$
$kT=\frac{H}{2\pi}$ and $A=\frac{4\pi}{H^{2}}$, respectively. Inserting these
physical quantities into Eq.(9), we obtain
\begin{equation}
E=\frac{1}{HL_{p}^{2}}.
\end{equation}
This is an explicit relation between the energy enclosed by the
Hubble horizon and the Hubble parameter. The relation (10) implies
that the universe enclosed by the Hubble horizon has a negative specific
heat since $E\propto T^{-1}$. This is also the expected result even
in Newtonian gravitating systems\cite{key-43}.

There are two relevant important energy contents in the flat universe,
namely the Misner-Sharp energy and Komar energy. Thus a question
naturally arises as to whether the energy $E$ in the energy equipartition
law is the Misner-Sharp energy or Komar energy. Now let us investigate
the two energy conditions.

\subsection*{3.1 Misner-Sharp Energy}

The Misner-Sharp energy is\cite{key-44}
\begin{equation}
E=\int T_{\mu\nu}u^{\mu}u^{\nu}dV,
\end{equation}
where $u^{\mu}=\delta_{0}^{\mu}$ is the four velocity and $T_{\mu\nu}=(\rho,\: p+p_{e},\: p+p_{e},\: p+p_{e})$
is the energy-momentum tensor. Then the energy inside the Hubble horizon
is
\begin{equation}
E=\int_{0}^{H^{-1}}4\pi r^{2}\rho dr=\frac{4\pi\rho}{3H^{3}},
\end{equation}
where we have assumed that the energy density $\rho$ is homogeneous.
Comparing Eq.(10) with Eq.(12), we get
\begin{equation}
\rho=\frac{3H^{2}}{4\pi L_{p}^{2}}.
\end{equation}
Thus we obtain one of the evolution equation of the universe from the
energy equipartition law. The equipartition law is an
equation of state while Eq.(13) can be deemed as a dynamic evolution equation of
the universe. So the evolution of the universe can be derived from
an equation of state, namely the equipartition law.

\subsection*{3.2 Komar Energy}

The Komar energy is defined as\cite{key-45,key-46}
\begin{equation}
E=\int(2T_{\mu\nu}-Tg_{\mu\nu})u^{\mu}u^{\nu}dV,
\end{equation}
where $T$ is the trace of the energy-momentum source $T_{\mu\nu}$.
Then the energy inside the Hubble horizon can be reduced to
\begin{equation}
E=\int_{0}^{H^{-1}}4\pi r^{2}|\rho+3(p+p_{e})|dr=\frac{4\pi}{3H^{3}}|\rho+3(p+p_{e})|.
\end{equation}
Thus Eq.(10) changes to
\begin{equation}
|\rho+3(p+p_{e})|=\frac{3H^{2}}{4\pi L_{p}^{2}}.
\end{equation}
This shows that the change of Hubble parameter is related to the
creation pressure of particles when the Komar energy is chosen. It is
also interesting to note that Padmanabhan thought that the Komor energy
is the active gravitational mass-energy\cite{key-18} and derived
the standard Friedmann equation when the Komar energy is taken\cite{key-22,key-23}.
Therefore, we will study the evolution of the universe by using
the Komar energy as the energy of the universe enclosed by Hubble
horizon in this paper. Of course, we can also use the Misner-Sharp energy as
the energy of the universe and obtain the evolution laws of the universe
consistent with astronomical observations. But we will not discuss it here.

As usual,  the cosmic fluid is deemed as the ideal fluid whose equation of state is
\begin{equation}
p=\omega\rho,
\end{equation}
where $\omega$ is a constant.

Here we would like to argue the validity of Eq.(8) from the viewpoint of gravitational dynamics.
If the gravity can be quantized and has a minimum quantum of area with
the order of $L_{p}^{2}$\cite{key-42}, then the horizon with
the area $A$ can be divided into $N=\frac{A}{c_{1}L_{p}^{2}}$ cells where
$c_{1}$ is a numerical factor. Then we assume that there are $c_{2}$ microscopic states for
the every cell, that is, every cell has $c_{2}$ degrees of freedom. According
to the energy equipartition law $(6)$, we obtain the total energy
\begin{equation}
E^{'}=\frac{c_{2}}{2}NkT=\frac{c_{2}}{c_{1}}\frac{A}{2L_{p}^{2}}kT.
\end{equation}
From Eq.(18), we find that Eq.(9) can be recovered when the relation $\frac{c_{2}}{c_{1}}=1$ is chosen.
On the other hand, Eq.(18) can be reduced to $\rho=\frac{3H^{2}}{8\pi L_{p}^{2}}$
which is the standard Friedmann equation for the flat FRW spacetime
if we choose $\frac{c_{2}}{c_{1}}=\frac{1}{2}$ and the Misner-Sharp energy (12).
That is, the energy equipartition
law and Einstein equation are equivalent to some extent when we
take the Misner-Sharp energy as the energy of the universe and choose $\frac{c_{2}}{c_{1}}=\frac{1}{2}$.
Thus, the Misner-Sharp energy seems to be a good choice for the energy of the universe.
But we use the Komar energy as the energy of the universe in this paper.
The reasons are as follows. Firstly, the Komar energy is really the gravitational energy as shown by Padmanabhan.
Secondly, the accelerated expansion of the present universe can not be described by the Einstein equation without a
cosmological constant, but can be explained by a modified gravitational field equation.
On the other hand, Eq.(16) can be derived by a modified gravitational field equation.
Thirdly, Eq.(8) rather than the Einstein equation is the ansatz from the perspective of thermodynamics.
The above discussions are only to show the validity of Eq.(8) from the viewpoint of gravitational dynamics
and do not mean that only the Minser-Sharp energy can explain the validity of Eq.(8).
The reason we use the Misner-Sharp energy to argue the validity of Eq.(8) is that
it is convenient to compare with the standard evolution equation derived from the general relativity.
Indeed, the modified Friedmann equation can be obtained when the Komar energy is chosen.

\section*{4. Physical process of the evolution of the universe}

In this section, we will give two specific cases to illustrate the
validity of the present model. Furthermore, we will analyze the specific
physical process of the evolution of the universe from the two cases.

Assuming that the rate of particle creation has the following formula
\begin{equation}
\Gamma=\gamma\left(\frac{H}{H_{0}}\right)^{n-1}H=\alpha H^{n},
\end{equation}
where $\gamma$ is a positive constant of the order 1, $n$ is a non-negative
integer, $H_{0}$ is a quantity with the same dimension as $H$ and
$\alpha=\frac{\gamma}{H_{0}^{n-1}}$. The purpose of introducing $H_{0}$
is to ensure that the dimension of $\Gamma$ is consistent with that
of $H$. Combining Eq.(4), Eq.(5) and Eq.(17), we obtain the specific modified continuity equation
\begin{equation}
\dot{\rho}+3(1+\omega)H\rho\left(1-\frac{\alpha}{3}H^{n-1}\right)=0.
\end{equation}
Besides, the dynamic evolution equation of the universe (16) can be reduced
to
\begin{equation}
\mid\alpha(1+\omega)H^{n-1}-(1+3\omega)\mid\rho=\frac{3H^{2}}{4\pi L_{p}^{2}}.
\end{equation}
Thus Eq.(20) and Eq.(21) constitute the fundamental equations of the
evolution of the universe. As long as the two equations are combined,
some basic physical quantities of the evolution of the universe, such
as the scale factor $a(t)$ and energy density $\rho$, can be solved.

\subsection*{4.1 Case of $\Gamma/H=\alpha H$: evolution of the early universe}

A natural choice is $n=2$, namely $\Gamma/H=\alpha H$. According
to the astronomical observations and cosmological theories, we know that
the Hubble parameter $H$ has been decreasing since the very early
stage of the universe except for the period of inflation. In this case,
the particle creation rate is very high in the early universe and decreases rapidly over time, so
choosing such a particle creation rate to study the evolution of the early universe is reasonable. 
In this subsection, $\omega$ is taken as $\frac{1}{3}$ because the early universe is
dominated by the radiation.

Under the above choice, Eq.(20) and Eq.(21) are reduced to
\begin{equation}
\dot{\rho}+4H\rho-\frac{4\alpha}{3}H^{2}\rho=0
\end{equation}
and
\begin{equation}
\mid\frac{4\alpha}{3}H-2\mid\rho=\frac{3H^{2}}{4\pi L_{p}^{2}}.
\end{equation}
Combining the above equations, we obtain the result
\begin{equation}
(\alpha H-3)(-6H^{3}+4\alpha H^{4}-3H\dot{H})=0.
\end{equation}
This equation has two solutions
\begin{equation}
H=\alpha/3
\end{equation}
and
\begin{equation}
t=\frac{\alpha}{3}\ln\mid\frac{4\alpha H-6}{H}\mid+\frac{1}{2H},
\end{equation}
where the integration constant is chosen as 0.

The solution (25) implies that there exists an inflation solution
$a(t)\propto e^{\frac{\alpha t}{3}}$ in the early universe. On the
other hand, the evolution law of the universe is $H=\frac{1}{2t}$
when $t$ is large from Eq.(26), which can be reduced to the standard
evolution law $a(t)\propto t^{1/2}$ in the radiation dominated stage.
These solutions can be explained as follows. The universe starts from
an unstable de Sitter space $(a(t)\propto e^{\frac{\alpha t}{3}}$),
then evolves to the standard radiation phase $(a(t)\propto t^{1/2})$.
During the evolution of the early universe, the creation rate of the radiation
is always $\Gamma=\alpha H^{2}$.

\subsection*{4.2 Case of $\Gamma/H=\alpha$: evolution of the present and late
universe}

Another natural choice is $n=1$, namely $\Gamma/H=\alpha$. Under
this choice, Eq.(20) and Eq.(21) are reduced to
\begin{equation}
\dot{\rho}+(1+\omega)\left(3-\alpha\right)H\rho=0
\end{equation}
and
\begin{equation}
\mid(\alpha-1)+(\alpha-3)\omega\mid\rho=\frac{3H^{2}}{4\pi L_{p}^{2}}.
\end{equation}
From Eq.(27) and Eq.(28), we obtain the results that $\rho=constant$
and $H^{2}=8\pi L_{p}^{2}\rho/3$ if the parameter $\omega$
is taken as $-1$. The results are consistent with the evolution laws
of the universe when the universe is dominated by the vacuum energy
in the general relativity. On the other hand, the same results are obtained
if $\alpha=3$. The fact implies that the evolution law of the universe
is $a(t)\propto e^{Ht}$ and independent of the nature of matter when
the particle creation rate $\Gamma$ equals $3H$. Therefore, we
can describe the exponentially expanding de Sitter universe without
introducing the vacuum energy which is equivalent to the negative
pressure matter when $\alpha=3$.

When $\omega\neq-1$ and $\alpha\neq3$, combining Eq.(27) and Eq.(28),
we obtain
\begin{equation}
\dot{H}=-\frac{1}{\delta} H^{2}.
\end{equation}
Further, we obtain
\begin{equation}
a(t)=t^{\delta},
\end{equation}
where $\delta=\frac{2}{(1+\omega)(3-\alpha)}$. The parameter $\delta$ satisfies the
inequality $\delta>1$ because
the expansion of the universe is speeding up according to the astronomical
observations. If the universe is dominated by the pressureless matter at present,
namely $\omega=0$, then we can get the range of the parameter $\alpha$
which is $1<\alpha<3$. From this result, we can describe the accelerated
expansion of the universe without introducing the dark energy. For
example, the authors of Ref.\cite{key-47} pointed out that the rate
of expansion which is consistent with supernova observations is $a(t)=t^{2}$ at
present. Such a rate of expansion can be obtained when $\alpha=2$
for the present universe dominated by the pressureless matter. Hence, we can
describe the accelerated expansion of the current universe and the
evolution of the de Sitter universe at a late time under this choice.

So far we have analyzed the specific physical processes of the evolution
history of the universe from the cases of $\Gamma/H=\alpha H$ and
$\Gamma/H=\alpha$ based on the particle creation and holographic
equipartition. These results are consistent with the conclusions of
the analysis of $\Lambda$ dark energy model which is generally accepted.
Therefore, the current model can well explain the whole evolution
history of the universe without introducing dark energy.

\section*{5. Thermodynamical properties based on the holographic equipartition}

Now we start to investigate the thermodynamical properties of the universe based on the holographic equipartition
in this model. In order to see the energy equipartition,
we calculate the physical quantity
\begin{equation}
\frac{1}{2}\beta E=\frac{1}{2}\frac{2\pi}{kH}\frac{1}{HL_{p}^{2}}=\frac{\pi}{H^{2}L_{p}^{2}}.
\end{equation}
So the relation
\begin{equation}
S=\frac{1}{2}\beta E
\end{equation}
holds in this model since $S=\frac{A}{4L_{p}^{2}}=\frac{\pi}{H^{2}L_{p}^{2}}$.
This is an important relation of thermodynamics which has been shown
by Padmanabhan in the cases of the static spacetime in the general
relativity\cite{key-18} and a wider class of gravity theories like the Lanczos-Lovelock
gravity\cite{key-48}. This relation is indeed the energy equipartition
law in the bulk when the holographic equipartition $N_{sur}=N_{bulk}$
is satisfied since $E=\frac{2}{\beta}S=\frac{kT}{2}\frac{A}{L_{p}^{2}}=\frac{1}{2}N_{sur}kT=\frac{1}{2}N_{bulk}kT$.
So the ansatz that the holographic equipartition is always satisfied throughout the evolution of the universe
is consistent with the conclusion that the relation $S=\frac{1}{2}\beta E$ is the energy equipartition in the static spacetime.
Moreover, the application of the energy equipartition law is generalized to the dynamic spacetime.
In return, the validity of relation (32) also means that it is a reasonable ansatz that
the universe obeys always the holographic equipartition.

Besides, differentiating Eq.(10), we obtain the change of total energy enclosed
by the Hubble horizon during a time interval $dt$
\begin{equation}
dE=-\frac{\dot{H}}{H^{2}L_{p}^{2}}dt.
\end{equation}
On the other hand, we obtain
\begin{equation}
TdS=\frac{H}{2\pi}\left(-\frac{2\pi\dot{H}}{H^{3}L_{p}^{2}}\right)dt=-\frac{\dot{H}}{H^{2}L_{p}^{2}}dt,
\end{equation}
where the temperature $T=\frac{H}{2\pi}$ and the entropy-area
relation $S=\frac{A}{4L_{p}^{2}}$ are used. Comparing Eq.(33) with
Eq.(34), we obtain the relation
\begin{equation}
dE=TdS.
\end{equation}
This relation is actually the energy conservation relation
for the universe enclosed by the Hubble horizon. However, the first
law of thermodynamics is expressed as $-dE=TdS$ in Ref.\cite{key-49}
(In fact, the relation is also used in some references, for example, \cite{key-35,key-50,key-51,key-52,key-53,key-54}).
Let us explain the difference between the notion $dE$ used in Ref.\cite{key-49}
and that used in the present paper. In Ref.\cite{key-49}, $-dE$
is the amount of heat flux crossing the horizon during time $dt$. But
here $dE$ is the change of the energy inside the Hubble horizon.
Moreover, $-dE$ is defined by $-dE=4\pi R^{2}T_{\mu\nu}k^{\mu}k^{\nu}dt$
where $k^{\mu}$ is the future directed ingoing null vector field
in Ref.\cite{key-49}. But here $dE$ is mainly composed of two parts, one
of which is related to the particle creation and the other to the
particles across the Hubble horizon, so the change of the entropy
in this paper is mainly caused by the change of the number of particles. In addition, the energy $E$ in the bulk can be also considered as
the energy on the horizon due to the holographic equipartition. Further,
the entropy change on the horizon $dS$ can be expressed as $dS=dS_{i}+dS_{e}$, where $dS_{i}$ is
the irreversible entropy change caused by the particle creation and $dS_{e}$ is
the reversible entropy change caused by the particles across the Hubble horizon.
Therefore, the non-equilibrium thermodynamics accompanied with the entropy production
can be expressed as the form $(35)$  of the equilibrium thermodynamics on the Hubble horizon.

\section*{6. Conclusions}

In this paper, we analyze the evolution history of the universe
based on the particle creation and holographic
equipartition. We assume that the universe always obeys the holographic
equipartition $N_{bulk}=N_{sur}$ throughout the evolution of the
universe due to the particle creation. This is a reasonable assumption
because there exists a simple and explicit holographic correspondence
between the surface term and bulk term of the Lagrange of gravity.
Further, we obtain an evolution equation of the universe by using
the energy equipartition and taking the Komar energy as the energy
of gravity. In the early universe, we choose the creation rate of
the radiation $\Gamma=\alpha H^{2}$ and obtain two solutions $a(t)\propto e^{\alpha t/3}$
and $a(t)\propto t^{1/2}$. In the present and late universe, we obtain
two solutions $a(t)\propto t^{\delta}$ and $a(t)\propto e^{Ht}$
by choosing the creation rate of the pressureless matter $\Gamma=\alpha H$. These
solutions are in good agreement with astronomical observations.

Based on the above results, we think that the evolution history of the universe
can be explained as follows. The universe starts from an
unstable de Sitter universe ($a(t)\propto e^{\alpha t/3}$) and evolves into a standard radiation
stage ($a(t)\propto t^{1/2}$) due to the creation of radiation. Then
the pressureless matter starts to dominate the evolution of the universe
due to the expansion of the universe. In the present universe dominated
by the pressureless matter, the universe expands as the law $a(t)\propto t^{\delta}$.
So we can explain the current accelerated expansion as long as the
parameter $\delta$ is greater than 1. Finally, the universe evolves
to the de Sitter Universe $(a(t)\propto e^{Ht})$ in the late stage.
Although the mechanisms of the transition from $a(t)\propto e^{\alpha t/3}$
to $a(t)\propto t^{1/2}$ in the early universe and the transition
from $a(t)\propto t^{\delta}$ to $a(t)\propto e^{Ht}$ in the present
and late universe are not clear. For completeness, we also obtain and discuss the thermodynamic relation $dE=TdS$
and conclude that the relation $S=\frac{1}{2}\beta E$
is indeed the energy equipartition law in our scenario.

\section*{Acknowledgments}
This research was funded by Doctoral Foundation of Zunyi Normal University (Grant No. BS[2016]03),
Education Department Foundation of Guizhou Province (Grant No. QianjiaoheKYzi[2017]247), Major Research
Project for Innovative Group of Education Department of Guizhou Province (Grant No. KY[2018]028) and the
NNSF of China (Grants No. 11775187, No. 11847031, No. 11865018 and No. 11865019).

\end{document}